\documentclass[conference]{IEEEtran}
\IEEEoverridecommandlockouts

\usepackage{amsmath,amssymb,amsfonts}
\usepackage{algorithmic}
\usepackage{graphicx}
\usepackage{textcomp}
\usepackage{xcolor}
\def\BibTeX{{\rm B\kern-.05em{\sc i\kern-.025em b}\kern-.08em
    T\kern-.1667em\lower.7ex\hbox{E}\kern-.125emX}}
    
    \allowdisplaybreaks
\begin{document}

\title{Energy-Efficient Task Offloading for Vehicular Edge Computing: Joint Optimization of Offloading and Bit Allocation\\
}

\author{\IEEEauthorblockN{Youngsu Jang, Jinyeop Na, Seongah Jeong$^\dag$ and Joonhyuk Kang}
\IEEEauthorblockA{Korea Advanced Institute of Science and Technology (KAIST), Daejeon, Korea \\
Email: \{jangyoung30, wlsduq37\}@kaist.ac.kr, jhkang@ee.kaist.ac.kr}
\IEEEauthorblockA{$^\dag$Kyungpook National University, Daegu, Korea \\
Email: seongah@knu.ac.kr}
}

\maketitle

\begin{abstract}
With the rapid development of vehicular networks, various applications that require high computation resources have emerged. To efficiently execute these applications, vehicular edge computing (VEC) can be employed. VEC offloads the computation tasks to the VEC node, i.e., the road side unit (RSU), which improves vehicular service and reduces energy consumption of the vehicle. However, communication environment is time-varying due to the movement of the vehicle, so that finding the optimal offloading parameters is still an open problem. Therefore, it is necessary to investigate an optimal offloading strategy for effective energy savings in energy-limited vehicles. In this paper, we consider the changes of communication environment due to various speeds of vehicles, which are not considered in previous studies. Then, we jointly optimize the offloading proportion and uplink/computation/downlink bit allocation of multiple vehicles, for the purpose of minimizing the total energy consumption of the vehicles under the delay constraint. Numerical results demonstrate that the proposed energy-efficient offloading strategy significantly reduces the total energy consumption.
\end{abstract}

\begin{IEEEkeywords}
Vehicular edge computing, energy efficiency, task offloading, bit allocation, vehicular networks.
\end{IEEEkeywords}

\section{Introduction}
The emergence of the Internet of things (IoT) revolutionizes the computing era. IoT means a large network of interconnected devices, and IoT can be useful for various applications such as health care, aviation, social networking, transportation, and traffic control \cite{1}. With the ever-increasing number of vehicles, the Internet of vehicles (IoV), which is typical application of IoT in the transportation area, has received much attention lately \cite{2}. Furthermore, various vehicular applications such as autonomous driving, video streaming, speech recognition, and in-vehicle entertainment are expected to be implemented in vehicles \cite{3}. However, these applications require not only strict delay constraints, but also enormous computation resources to process large volumes of workload data. Since existing cloud servers are located far from the vehicle, offloading tasks to the cloud server causes latency problems due to the capacity-constrained backhaul links \cite{4}. To tackle this problem, vehicular edge computing (VEC), which provides the service at the network edge near the vehicle, has been proposed as a solution \cite{5}. \\
\indent VEC is expected to support computation sensitive and delay sensitive applications effectively by offloading workloads to the VEC nodes that are close to the vehicle. Traditionally, all applications need to be processed locally on the vehicle, which causes enormous energy consumption, not desirable for energy-limited vehicles such as electric cars. In VEC, the vehicle can offload the tasks to adjacent VEC node, which has higher computational capability and abundant energy supply \cite{2}. Thus, VEC can reduces the energy consumption of vehicle and increases the battery endurance time. However, task offloading does not always guarantee low energy consumption due to the energy consumption for communication. Therefore, it is necessary to study the energy-efficient offloading strategies. \\
\indent Recently, there are some works that investigate offloading strategies in VEC. In \cite{2}, an offloading scheme is designed to enhance the transmission efficiency under the considerations of the time consumption of the task execution and the mobility of the vehicles. The authors in \cite{4} develop a mobile edge computing (MEC) offloading framework to support various applications in smart city scenarios. Also, a VEC framework named autonomous vehicular edge (AVE) is proposed in \cite{6} to increase the computational capabilities of vehicles in a decentralized manner. However, these works mainly solve the workload offloading problem from the perspective of minimizing latency, and have not considered the energy saving problems. Although, there are a few studies that consider the energy-efficient task offloading.  An energy-efficient VEC framework is proposed in \cite{3} to support in-vehicle user equipments (UEs) which have limited battery capacity. In \cite{7}, an energy-efficient workload offloading problem is studied, and a low-complexity distributed solution based on consensus alternating direction method of multipliers (ADMM) is proposed. The authors in \cite{8} investigate the average transmission power minimization problem under the tasks quality of service (QoS) requirement. Nevertheless, \cite{3} and \cite{7} are mainly target on in-vehicle UEs, and \cite{8} have not considered the multiple vehicle case. \\
\indent In this paper, we consider the VEC system that offloads workload to nearby road side unit (RSU) with computation server, when multiple vehicles arrive at different times on a unidirectional road with multiple lanes. The problem that minimizes the total energy consumption of vehicles is formulated as a joint bit allocation and offloading proportion problem. The main contributions of this paper are as follows:
\begin{itemize}
\item We formulate the energy-efficient task offloading problem to minimize the total energy consumption of all vehicles, under the consideration of the task deadline, computation and communication energy consumption, multiple access, and time-varying channel state. In addition, we further consider the offloading across multiple RSUs, various velocities of vehicles, and multiple lanes, that have been ignored in previous works \cite{3}, \cite{7}, and \cite{8}. 
\item We deal with the optimization problems for two different cases in the offloading process. One is the complete offloading case where the vehicles offload all their tasks. The other one is the partial offloading case where the vehicles offload a fraction of their tasks. Also, we provide the optimal solution for both cases. Simulation results demonstrate the effectiveness of the proposed offloading strategy.
\end{itemize}
\indent \indent The rest of the paper is organized as follows. First, the system model including the computation energy model and the communication energy model are introduced in Section II. Section III describes the formulation of optimization problem. The simulation results and analysis are depicted in Section IV. Finally, we conclude the paper in Section V.

\section{System Model}
\begin{figure}[!t]
\centerline{\includegraphics[width=3.4in]{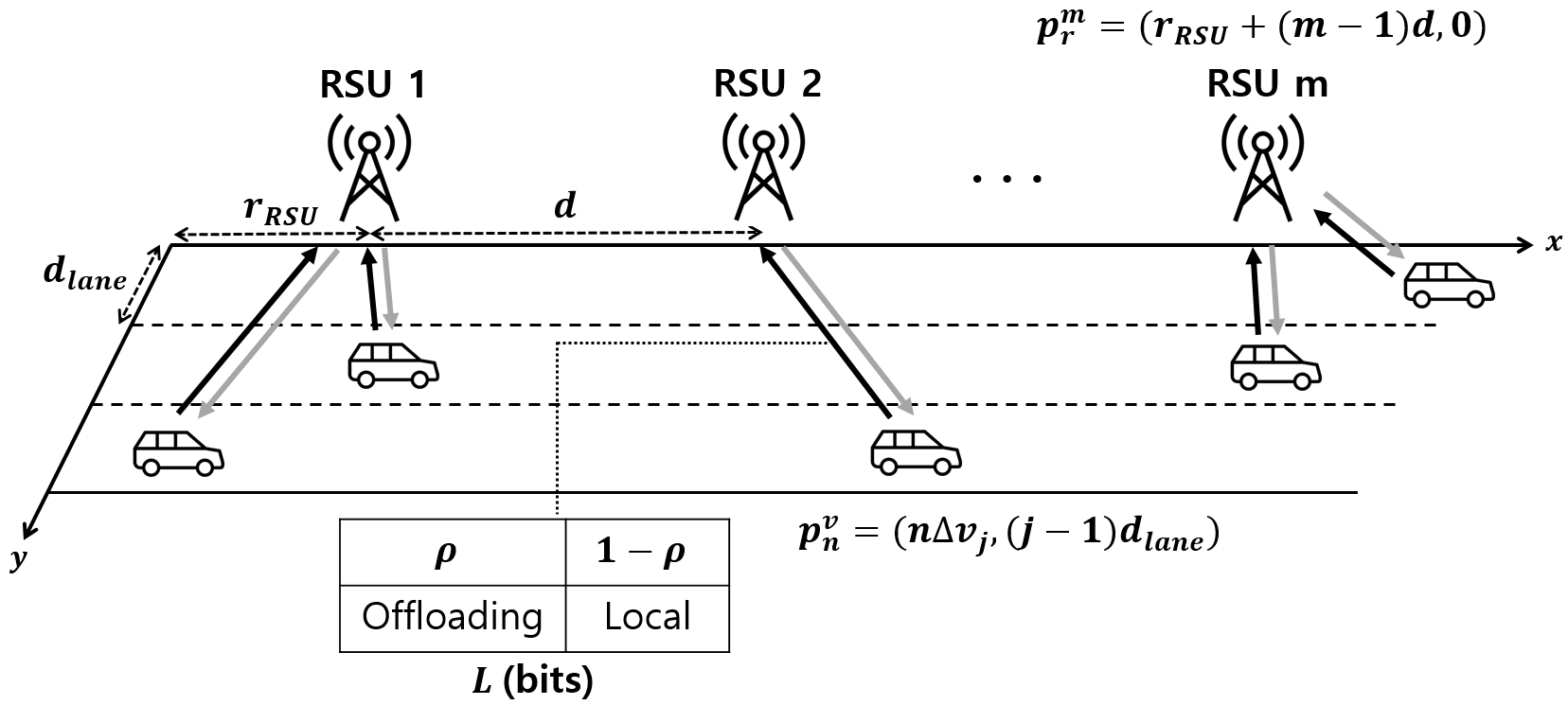}}
\caption{Illustration of the task offloading in VEC}
\label{fig1}
\end{figure}

In this paper, we consider a VEC system that includes $K$ vehicles and $M$ RSUs as shown in Fig.~\ref{fig1}. The RSUs are deployed along the unidirectional road with $J$ lanes, the distance between neighboring RSUs is $d$, and the coverage radius of each RSU is $r_ {\text{RSU}}$. Define $\mathcal{M}=\{1,\ldots,M\}$ as the set of RSUs. The location of RSU $m$ is calculated as
\begin{equation}
\mathbf{p}_m^r=(r_{\text{RSU}}+(m-1)d,0),\;\;m\in\mathcal{M}. \label{eq1}
\end{equation}
We assume that $K$ vehicles arrive at the first RSU's coverage edge in different time $t_k\in\{t_1,\ldots,t_K\}$. The set of vehicles is defined as $\mathcal{K}=\{1,\ldots,K\}$. By further assuming that the vehicles in the same lane have the same velocity, the velocity of the vehicles in each lane is $v_j\in\{v_1,\ldots,v_J\}$ \cite{9}.\\
\indent The offloading process is described in the following steps. First, vehicle $k$ transmits the input data to the closest RSU, i.e., uplink transmission. Next, the RSU computes the received data, and finally, the RSU transmits the output data to vehicle $k$, i.e., downlink transmission. We assume frequency division duplex (FDD) that equal bandwidth $B$ is allocated for uplink and downlink. Thus, there is no interference between uplink and downlink communication. The time horizon $T$ is equally divided into $N$ frames, and each frame duration is $\Delta$, i.e., $T=N\Delta$. The frame duration $\Delta$ is sufficiently small so that the vehicle's location is approximately constant within each frame. Thus, the position of the vehicle in the $j$th lane at the $n$th frame is expressed as
\begin{eqnarray} 
\mathbf{p}_n^v=(n\Delta v_j,(j-1)d_{\text{lane}}),\qquad \label{eq2} \\ 
 j=1,\ldots,J\;\textrm{ and }n\in\mathcal{N}=\{1,\ldots,N\}, \nonumber
\end{eqnarray}
where $d_{\text{lane}}$ denotes the lane width. For the access scheme, we consider an orthogonal access that each vehicle has one time slot with duration $\delta=\Delta/K$ and communicates within its time slot.\\
\indent For the sake of simplicity, we consider the communication channels between vehicles and RSUs which are dominated by line-of-sight (LoS) link \cite{10}. Doppler effect due to the vehicle's movement is assumed to be perfectly compensated by the receivers. In each frame, the vehicle communicates with the nearest RSU. Thus, the channel gain between the vehicle $k$ and the adjacent RSU at the $n$th frame is given by 
\begin{equation}
h_{k,n}(\mathbf{p}_n^v)=\frac{h_0}{\lVert\mathbf{p}_n^v-\mathbf{p}_{m_{\min}}^r\rVert^2+H^2}, \label{eq3}
\end{equation}
where $\Vert.\Vert$ means the norm 2 function, $m_{\min}\in\mathcal{M}$ represents the index of the closest RSU, $H$ denotes the height of the RSU, and $h_0$ is the received power at the reference distance $d=1$m for a transmission power of $1$W. We assume that the channel noise is additive white Gaussian with zero mean and power spectral density $N_0$ [dBm/Hz].\\
\indent The task of the vehicle $k\in\mathcal{K}$ is described by the number $L_k$ of input bits, the number $C_k$ of CPU cycles per input bit for computation, and the number $\kappa_k$ of output bits produced by computation per input bit. In most cases, the size of input data is much larger than the output data, i.e., $\kappa_k<1$.  All tasks must be computed within the deadline $T$.
\subsection{Computation energy model}
When the CPU is operated at frequency $f^i$, the computation energy consumption to execute the application of vehicle $k$ with $l$ input bits is obtained by \cite{11}, \cite{12}
\begin{equation}
E_k^i(l,f^i)=\gamma^i C_kl(f^i)^2, \label{eq4}
\end{equation}
where $i=v$ for the vehicle and $i=r$ for the RSU, respectively. Here, $f^i$ [CPU cycles/s] is the operating frequency of each processor, and $\gamma^i$ denotes the effective switched capacitance of each processor which is related to the chip architecture.
\subsection{Communication energy model}
When vehicle $k$ transmits $L_{k,n}^q$ bits during the slot duration $\delta$, the following equation is obtained by Shannon theory.
\begin{equation}
B\delta \log_2\Bigg(1+\frac{E_{k,n}^q(L_{k,n}^q,\mathbf{p}_n^v)h_{k,n}(\mathbf{p}_n^v)}{N_0B\delta}\Bigg)=L_{k,n}^q, \label{eq5}
\end{equation}
where $q=u$ for uplink while $q=d$ for downlink. From \eqref{eq5}, the communication energy consumption of the vehicle $k$ at time slot $n$ is calculated as
\begin{equation}
E_{k,n}^q(L_{k,n}^q,\mathbf{p}_n^v)=\frac{N_0B\delta}{h_{k,n}(\mathbf{p}_n^v)}\bigg(2^{\frac{L_{k,n}^q}{B\delta}}-1\bigg). \label{eq6}
\end{equation}
From \eqref{eq6}, we can see that the communication energy consumption is related to the number of transmission bits and the channel condition affected by the communication distance.

\section{Minimize Total Energy Consumption of Vehicles}
In this section, we study the problem to minimize the total energy consumption of $K$ vehicles. First, the energy consumption when locally computing in  the vehicle is obtained as a benchmark for comparison. Then, without considering the offloading proportion, the optimal bit allocation for uplink, computation, and downlink is found under the maximum delay constraint of $T$. Finally, the joint optimization of bit allocation and offloading ratio is introduced.

\subsection{Energy consumption for local execution}
For reference, we consider the total energy consumption of overall vehicles when all applications are processed locally. To process $L_k$ bits within $T$ seconds, the CPU frequency of vehicle $k$ needs to be selected as
\begin{equation}
f_k^v=\frac{C_kL_k}{T}. \label{eq7}
\end{equation}
From \eqref{eq4}, the total energy consumption for local execution is obtained by
\begin{equation}
\sum_{k=1}^{K} E_k^v(L_k)=\sum_{k=1}^{K} \frac{\gamma_k^v{C_k^3}}{T^2}L_k^3, \label{eq8}
\end{equation}
where $\gamma_k^v$ is the effective switched capacitance of the vehicle $k$'s processor.

\subsection{Optimization for complete offloading}
In this section, we study the optimal bit allocation that minimize the total energy consumption of all vehicles. It is assumed that all input data for each vehicle is offloaded to the RSU so that the data is not processed locally in the vehicle. At the $n$th frame, we denote $L_{k,n}^u$ as the number of uplink bits transmitted from the vehicle $k$ to the RSU, $l_{k,n}^c$ as the number of bits computed for the task of the vehicle $k$ at the RSU, and $L_{k,n}^d$ as the number of downlink bits transmitted from the RSU to the vehicle $k$. Also, we define $\tilde{\mathcal{N}}=\{1,\ldots,N-2\}$. \\
\indent We analyze the offloading process in frame-by-frame manner. This method is commonly used to handle the offloading process \cite{13}. For example, vehicle $k$ transmits $L_{k,1}^u$ bits to the RSU at the first frame and there is no computation and no downlink transmission, i.e., $l_{k,1}^c=L_{k,1}^d=0$. At the second frame, vehicle $k$ transmits $L_{k,2}^u$ bits to the RSU, the RSU computes $l_{k,2}^c \le L_{k,1}^u$ bits, and there is no downlink transmission, i.e., $L_{k,2}^d=0$. At the third frame, vehicle $k$ transmits $L_{k,3}^u$ bits to the RSU, the RSU computes $l_{k,3}^c \le L_{k,1}^u+L_{k,2}^u-l_{k,2}^c$ bits, and the RSU transmits $L_{k,3}^d \le \kappa_kl_{k,2}^c$ bits to vehicle $k$. This process is continued until the $N$th frame. The optimization problem is formulated as follows:
\begin{subequations} 
\label{eq9}\begin{align}
& \underset{\{L_{k,n}^u\},\{l_{k,n}^c\},\{L_{k,n}^d\}}{\text{minimize}} \sum_{k=1}^{K}\sum_{n=1}^{N-2}E_{k,n}^u(L_{k,n}^u) \label{eq9a} \\
& \text{s.t.}\; \sum_{i=1}^{n}l_{k,i+1}^c \leq \sum_{i=1}^{n}L_{k,i}^u, \;\textrm{for}\: k \in \mathcal{K} \:\textrm{and}\: n \in \tilde{\mathcal{N}}, \label{eq9b} \\
& \;\;\;\;\;\sum_{i=1}^{n}L_{k,i+2}^d \leq \kappa_k\sum_{i=1}^{n}l_{k,i+1}^c, \;\textrm{for}\: k \in \mathcal{K} \:\textrm{and}\: n \in \tilde{\mathcal{N}}, \label{eq9c} \\
& \;\;\;\;\; L_{k,n}^u \leq L^u_{\max}, \;\textrm{for}\: k \in \mathcal{K} \:\textrm{and}\: n \in \tilde{\mathcal{N}}, \label{eq9d} \\
& \;\;\;\;\; L_{k,n+2}^d \leq L^d_{\max}, \; \textrm{for}\: k \in \mathcal{K} \:\textrm{and}\: n \in \tilde{\mathcal{N}}, \label{eq9e} \\
& \;\;\;\;\; \sum_{k\in\mathcal{A}}L_{k,n}^u \leq L_{\max}, \;\textrm{for}\: k \in \mathcal{A} \: \textrm{and}\: n \in \tilde{\mathcal{N}}, \label{eq9f} \\
& \;\;\;\;\; L_{k,n}^u,\:l_{k,n}^c,\:L_{k,n}^d \geq 0, \; \textrm{for}\: k \in \mathcal{K} \:\textrm{and}\: n \in \mathcal{N}, \label{eq9g} \\
& \;\;\;\;\; \sum_{n=1}^{N-2}L_{k,n}^u = L_k, \; \textrm{for}\: k \in \mathcal{K}, \label{eq9h} \\
& \;\;\;\;\; \sum_{n=1}^{N-2}l_{k,n+1}^c = L_k, \; \textrm{for}\: k \in \mathcal{K}, \label{eq9i} \\
& \;\;\;\;\; \sum_{n=1}^{N-2}L_{k,n+2}^d = \kappa_k L_k, \; \textrm{for}\: k \in \mathcal{K}, \label{eq9j}
\end{align}
\end{subequations}
where $L_{\max}^u$ and $L_{\max}^d$ represent the maximum number of uplink and downlink bits that can be transmitted within slot duration, $\mathcal{A}$ is the set of vehicles that communicate with the same RSU at $n$th frame, and $L_{\max}$ is the limitation of total uplink bits transmitted to the same RSU at each frame. \\
\indent In problem \eqref{eq9}, the constraint \eqref{eq9b} guarantees that the number of bits processed at the $(n+1)$th frame by the RSU is no larger than the number of uplink bits transmitted from the vehicle in the previous $n$ frames. In a similar manner, the constraint \eqref{eq9c} guarantees that the number of downlink bits transmitted from the RSU at the $(n+2)$th frame is no larger than the number of output bits processed in the previous $(n+1)$ frames. Next, \eqref{eq9d} and \eqref{eq9e} are the constraints of uplink and downlink transmission bits at each frame. Also, \eqref{eq9f} is the constraint of the number of uplink bits received by one RSU at each frame, which is mainly due to the computation capability of the RSU. The constraint \eqref{eq9g} enforces non-negative bit allocations. Finally, the equality constraints \eqref{eq9h}-\eqref{eq9j} ensure that the given input bits are completely processed via offloading within the deadline $T$.\\
\indent The problem \eqref{eq9} is convex, since the objective function \eqref{eq9a} is the sum of convex exponential functions and the constraints \eqref{eq9b}-\eqref{eq9j} are linear. Thus, we can solve this problem using standard convex optimization solver such as CVX MOSEK \cite{14}.

\subsection{Optimization for partial offloading}
In order to obtain the additional energy savings, we consider not only the bit allocation, but also the offloading proportion. Assume that vehicle $k$ offloads the ratio $\rho_k$ of the input bits to the VEC node and locally computes the remaining portion of the input bits. In this case, the energy consumption of the vehicle is the sum of the communication energy consumed in offloading process and computation energy consumed in local execution. Thus, the joint optimization problem of bit allocation and offloading proportion is as follows:
\begin{subequations} \label{eq10}
\begin{align}
& \underset{\{L_{k,n}^u\},\{l_{k,n}^c\},\{L_{k,n}^d\},\{\rho_k\}}{\text{minimize}} \sum_{k=1}^{K}\sum_{n=1}^{N-2}E_{k,n}^u(L_{k,n}^u) &\qquad\qquad \nonumber \\
&\qquad \qquad\qquad +\sum_{k=1}^{K}E_k^v((1-\rho_k)L_k) \label{eq10a}  \\
& \text{s.t.} \;\sum_{n=1}^{N-2}L_{k,n}^u = \rho_kL_k,  \; \textrm{for}\: k \in \mathcal{K}, \label{eq10b} \\
& \;\;\;\;\; \sum_{n=1}^{N-2}l_{k,n+1}^c = \rho_kL_k,  \; \textrm{for}\: k \in \mathcal{K}, \label{eq10c} \\
& \;\;\;\;\; \sum_{n=1}^{N-2}L_{k,n+2}^d = \kappa_k \rho_kL_k,  \; \textrm{for}\: k \in \mathcal{K}, \label{eq10d} \\
& \;\;\;\;\; 0 \leq \rho_k \leq 1,  \; \textrm{for}\: k \in \mathcal{K}, \label{eq10e} \\
& \;\;\;\;\; \eqref{eq9b}-\eqref{eq9g}, \label{eq10f} 
\end{align}
\end{subequations}
where the equality constraints \eqref{eq10b}-\eqref{eq10d} ensure the completion of total offloaded data, and the inequality constraint \eqref{eq10e} denotes the offloading ratio constraint.\\
\indent The objective function \eqref{eq10a} is the sum of convex exponential functions and convex power functions, and the constraints \eqref{eq10b}-\eqref{eq10f} are linear. Hence, the problem \eqref{eq10} is convex, which can be solved by CVX MOSEK.

\section{Numerical Results}
\begin{figure}[!t]
\centerline{\includegraphics[width=3.8in]{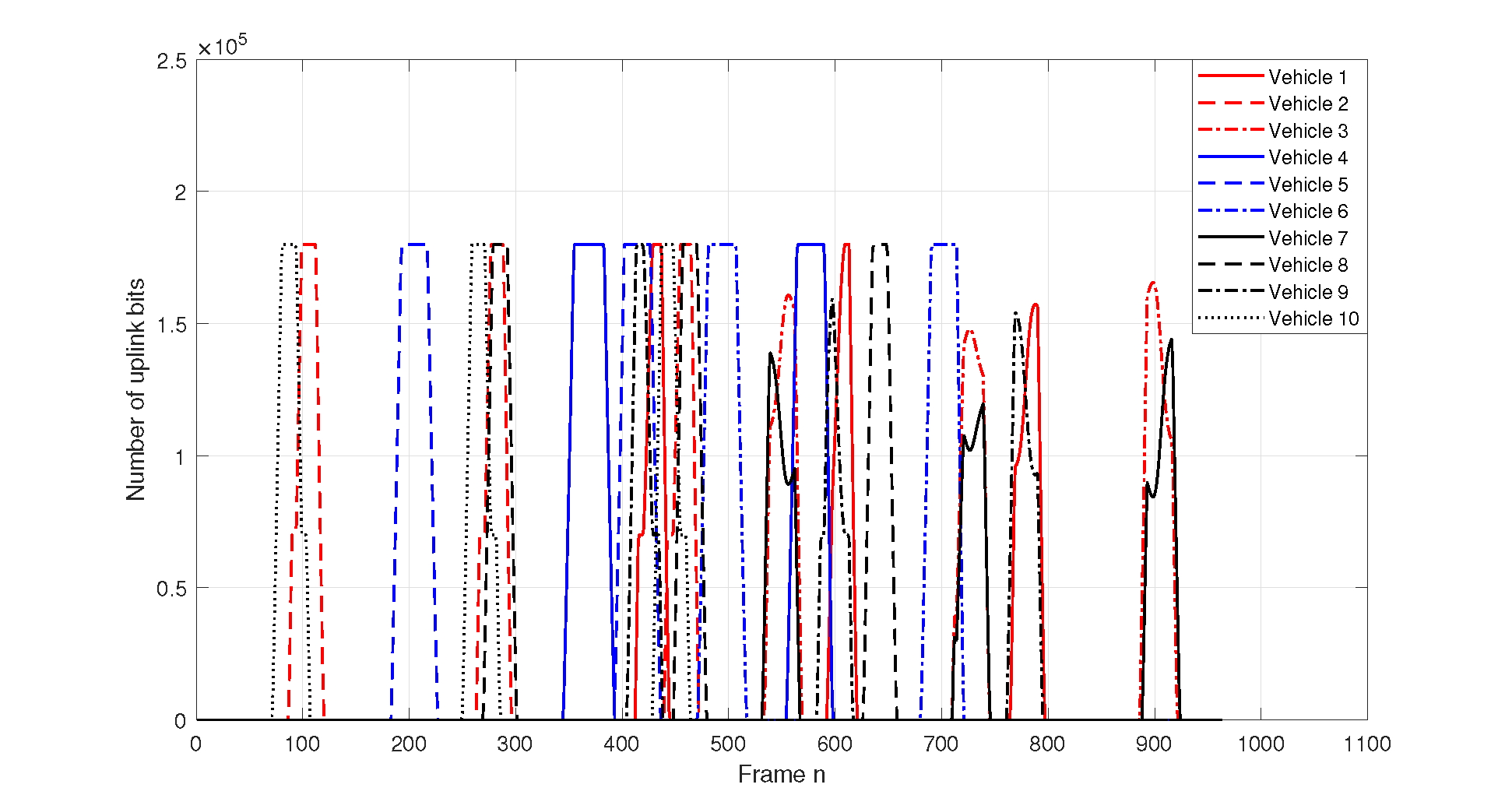}}
\caption{Optimized uplink bit allocation at each frame under $T=20$s}
\label{fig2}
\end{figure}

\begin{figure}[t]
\centerline{\includegraphics[width=3in]{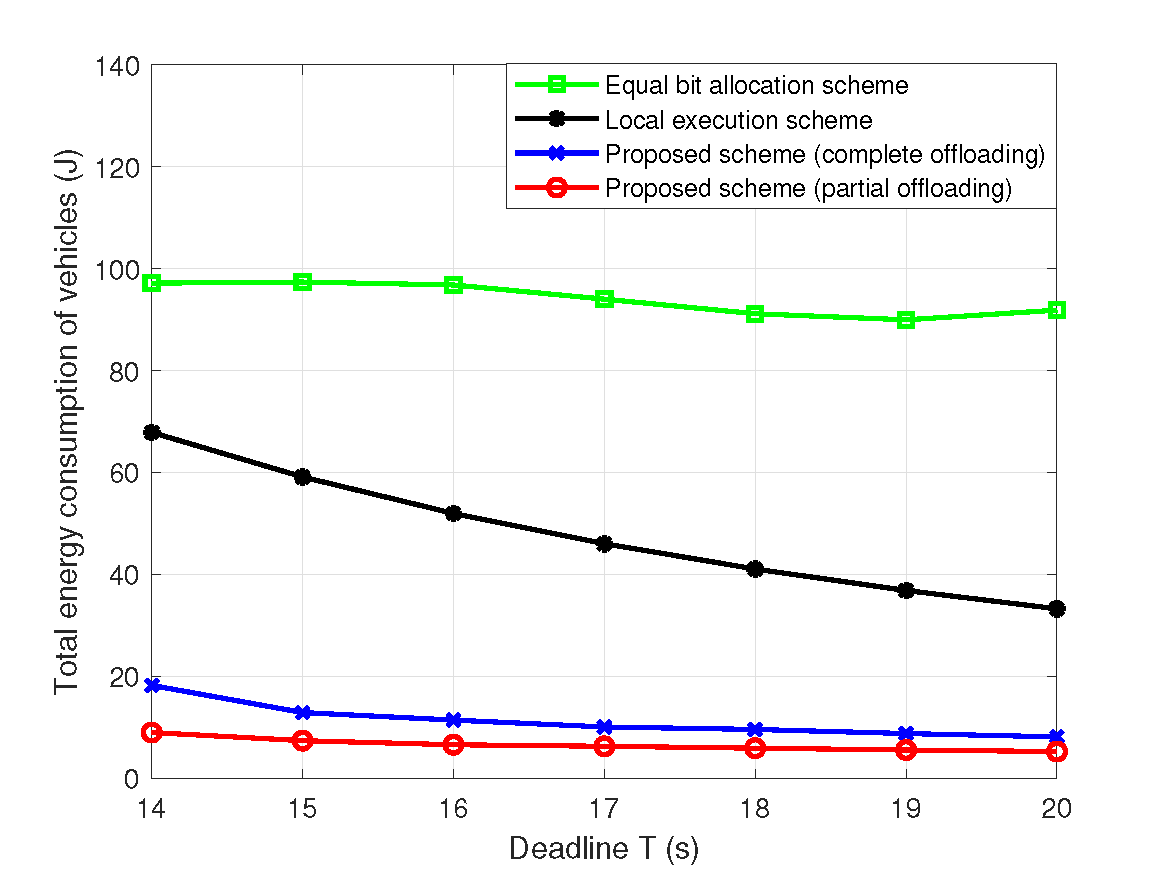}}
\caption{Total energy consumption of the vehicles versus different deadline $T$}
\label{fig3}
\end{figure}

\begin{figure}[t]
\centerline{\includegraphics[width=3in]{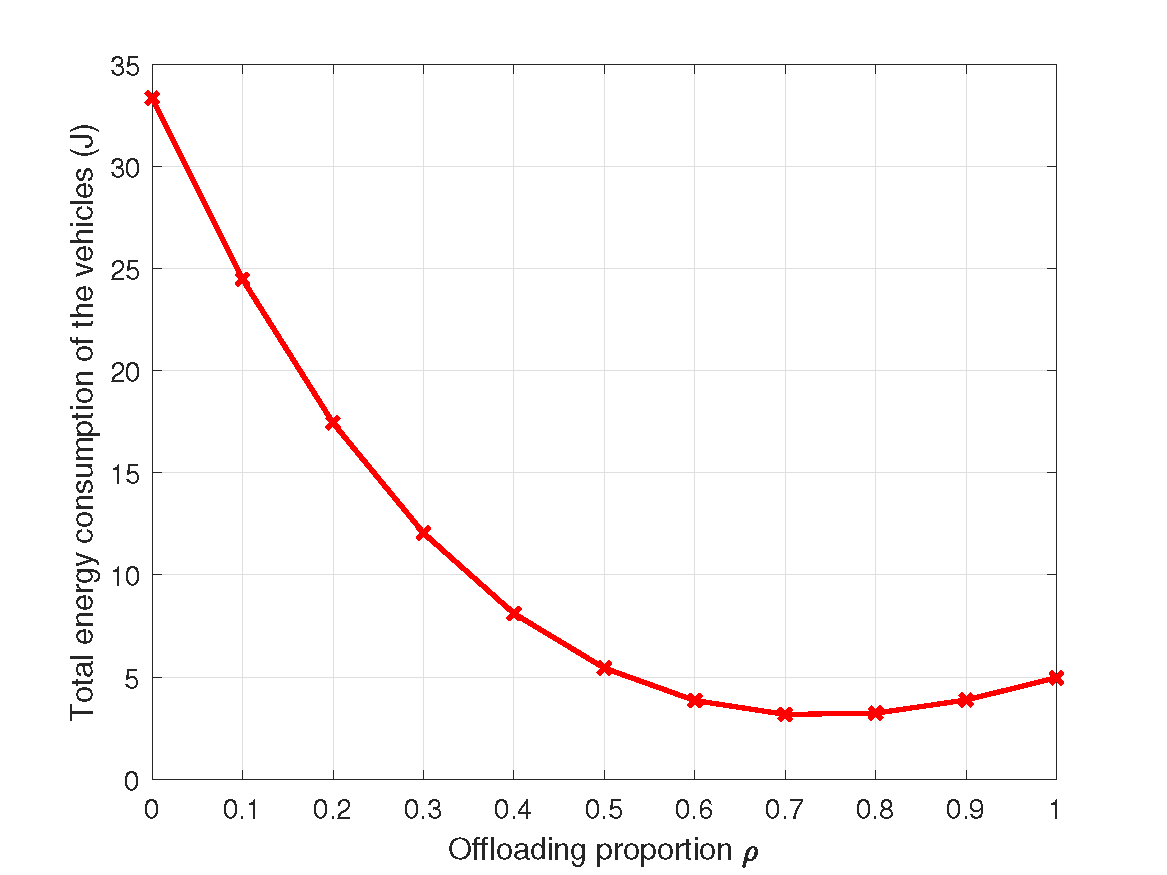}}
\caption{Total energy consumption of the vehicles with respect to the offloading proportion when all vehicles have equal offloading proportion}
\label{fig4}
\end{figure}

In this section, we provide simulation results to evaluate the performance of our proposed offloading design. In the simulations, we consider a system with the number of vehicles $K=10$ and the number of lanes $J=3$. The number of RSUs $M$ is determined by $M=\lceil (v_{\max}T-r_{\text{RSU}})/d+1/2\rceil$, where $v_{\max}$ is the velocity of the fastest vehicle and $\lceil . \rceil$ means the ceiling function. We assume that all vehicles have same deadline $T$ and the vehicles are arrived at the first RSU's coverage edge in randomly. The remaining parameters are summarized in Table \ref{table1}. We consider the following two schemes for reference. First, local execution scheme that all applications are processed locally. Second, we consider the equal bit allocation scheme. With this scheme, the same number of bits is transmitted in uplink and downlink in each frame, and the same number of bits is computed at the RSU at each frame, i.e., $L_{k,n}^u=l_{k,n+1}^c=L_k/(N-2)$ and $L_{k,n+2}^d=\kappa_kL_k/(N-2)$ for $k\in\mathcal{K}$  and $n=1,\ldots,N-2$. \\
\indent Fig.~\ref{fig2} represents the optimal uplink bit allocation at each frame. The red, blue, and black lines mean the vehicles in lane 1, 2, and 3, respectively. We set the deadline $T=20$s, but the proposed scheme can be applied even if the deadline is shorter. As shown in Fig.~\ref{fig2}, we can observe that the vehicles need to offload more data when they are closer to the RSU, to reduce the communication energy consumption. Additionally, when two or more vehicles communicate with the same RSU, some vehicles cannot offload the maximum number of bits due to the limited computation capability of the RSU, as described in constraint \eqref{eq9f}. In this case, the vehicle in the better channel condition transmits more bits. If the channel conditions of the vehicles are similar, the vehicle which has more total input bits, i.e., $\rho_kL_k$, offloads more bits. \\
\begin{table}[!t]
\renewcommand{\arraystretch}{1.2}
\caption{Simulation Parameters}
\begin{center}
\begin{tabular}{ccc}
\hline\hline
Parameter & Definition & Value \\
\hline
$r_{\text{RSU}}$ & coverage radius of RSU & 100m \\
$d$ & distance between neighboring RSUs & 200m \\
$d_{\text{lane}}$ & lane width & 3.5m \\
$H$ & height of RSU & 10m \\
$B$ & bandwidth & 40MHz \\
$L_k$ & number of input bits & 10 - 25Mbits \\
$L_{\max}^u$ & uplink constraint & 180Kbits \\
$L_{\max}^d$ & downlink constraint & 140Kbits \\
$L_{\max}$ & multiple access constraint & 250Kbits \\
$C_k$ & number of CPU cycles per bit & 1550.7 \;$\textrm{\cite{11}, \cite{12}}$ \\
$\gamma^r, \gamma_k^v$ & switch capacitance constant & $10^{-28} \;\textrm{\cite{11}, \cite{12}}$ \\
$\kappa_k$ & number of output bits per input bit & 0.5 \\
$N_0$ & noise spectrum density & -174dBm/Hz \\
$h_0/N_0B$ & reference SNR & 20dB \\
$v_j$ & velocity of vehicle & 20 - 30m/s \\
$\delta$ & time slot duration & 4ms \\
$\Delta$ & frame duration & 40ms \cite{15} \\
$t_k$ & arrival time of vehicle & 0 - 20s \\
\hline\hline
\end{tabular}
\end{center}
\label{table1}
\end{table}
In Fig.~\ref{fig3}, we compare the total energy consumption of each scheme as a function of the deadline $T$. From Fig.~\ref{fig3}, we can see that the energy savings of proposed offloading scheme becomes more effective compared to the benchmarks if the deadline $T$ becomes stricter. Also, we can observe that the total energy consumption of the equal bit allocation scheme is larger than the local execution scheme. This means that task offloading does not always guarantee low energy consumption due to the communication energy. In other words, the vehicles continue to transmit the input bits to the RSU when they are located in far from the RSU, so it causes a case where they consume more energy than when they process locally. Furthermore, it is seen that the proposed optimal bit allocation significantly reduces the total energy consumption, and the joint optimization of bit allocation and offloading ratio achieves additional energy savings. \\
\indent Fig.~\ref{fig4} shows the total energy consumption of all vehicles as a function of offloading proportion. In this simulation, we assume that all vehicles offload their task with equal offloading proportion $\rho$. In Fig.~\ref{fig4}, we can observe that the local execution, i.e., $\rho=0$, causes enormous energy consumption and offloading about 70\% of the task is more effective than offloading all the task, i.e., $\rho=1$. The reason for this is that if the vehicle moves beyond a certain distance away from the RSU, local execution consumes less energy than when data offloading is applied. Thus, in order to minimize the energy consumption, some portion of the task needs to be processed locally.

\section{Conclusion}
In this paper, we studied the energy-efficient task offloading strategy with the purpose of minimizing the total energy consumption of overall vehicles. First, we investigated the problem of finding the optimal uplink/computation/downlink bit allocation of multiple vehicles, when the vehicles offload all the input bits without considering the offloading proportion. We then formulated the joint optimization problem involving the offloading proportion of multiple vehicles. Since both optimization problems are convex, optimal bit allocation and offloading proportion are obtained by using CVX MOSEK. Numerical results demonstrate that the proposed joint optimization of bit allocation and offloading ratio achieves significant energy savings compared to the benchmarks.

\end{document}